\documentclass[twocolumn,english,aps,prl]{revtex4}
\usepackage[T1]{fontenc}
\usepackage[latin9]{inputenc}
\usepackage{babel}
\usepackage{float}
\usepackage{mathrsfs}
\usepackage{amsmath}
\usepackage{amssymb}
\usepackage{graphicx}
\usepackage{esint}

\makeatletter
\@ifundefined{textcolor}{}
{%
 \definecolor{BLACK}{gray}{0}
 \definecolor{WHITE}{gray}{1}
 \definecolor{RED}{rgb}{1,0,0}
 \definecolor{GREEN}{rgb}{0,1,0}
 \definecolor{BLUE}{rgb}{0,0,1}
 \definecolor{CYAN}{cmyk}{1,0,0,0}
 \definecolor{MAGENTA}{cmyk}{0,1,0,0}
 \definecolor{YELLOW}{cmyk}{0,0,1,0}
}

\makeatother

\begin{document}

\preprint{This line only printed with preprint option}

\title{Local Dynamics in Trained Recurrent Neural Networks}

\author{Alexander Rivkind}

\email{arivkind@tx.technion.ac.il}

\selectlanguage{english}%

\affiliation{Faculty of Medicine, Technion\textendash{}Israel Institute of Technology,
Haifa 32000, Israel}

\affiliation{Network Biology Research Labratories, Technion - Israel Institute
of Technology, Haifa 32000, Israel }

\author{Omri Barak}

\email{omri.barak@gmail.com}

\selectlanguage{english}%

\affiliation{Faculty of Medicine, Technion\textendash{}Israel Institute of Technology,
Haifa 32000, Israel}

\affiliation{Network Biology Research Labratories, Technion - Israel Institute
of Technology, Haifa 32000, Israel }
\begin{abstract}
Learning a task induces connectivity changes in neural circuits, thereby
changing their dynamics. To elucidate task related neural dynamics
we study trained Recurrent Neural Networks. We develop a Mean Field
Theory for Reservoir Computing networks trained to have multiple fixed
point attractors. Our main result is that the dynamics of the network's
output in the vicinity of attractors is governed by a low order linear
Ordinary Differential Equation. Stability of the resulting ODE can
be assessed, predicting training success or failure. As a consequence,
networks of Rectified Linear (RLU) and of sigmoidal nonlinearities
are shown to have diametrically different properties when it comes
to learning attractors. Furthermore, a characteristic time constant,
which remains finite at the edge of chaos, offers an explanation of
the network's output robustness in the presence of variability of
the internal neural dynamics. Finally, the proposed theory predicts\textbf{
}state dependent frequency selectivity in network response. 
\end{abstract}
\maketitle

\noindent Task learning is considered the \emph{raison d'etre} of
recurrent neural networks (RNN), studied in the context of neuroscience
and machine learning \cite{mante2013context,lecun2015deep}. Yet,
theoretical understanding of trained RNN dynamics is lacking, with
most of the existing physics literature addressing either random networks,
designed networks (\cite{hopfield1982neural,gardner1988space} and
\cite{ben1995theory}) or designed control setting \cite{Popovych2005Desynchronization,pyragas1992continuous,ott1990controlling}.

\noindent In this Letter, we advance a theory of trained RNN dynamics.
We consider an initially random, chaotic network whose output is trained
to produce several target values, and then fed back to the network,
yielding multiple fixed point attractors. This setting underlies complex
tasks that were analyzed phenomenologically using rate models \cite{mante2013context,Carnevale2015,sussillo2013opening},
and are the subjects of attempts \cite{abbott2016building} to extend
to more realistic task performing networks \cite{neymotin2013reinforcement}.
Using mean field analysis, we derive the effect of training on the
output dynamics in the vicinity of the training targets. Stability
is then assessed, showing that training success depends on the network's
nonlinearity. Next, we show that multiple training targets can lead
to state specific frequency selectivity, as observed in task adapted
biological neuronal circuits \cite{buzsaki2006rhythms,siegel2015cortical}.
Finally, the settling time of an output of a perturbed RNN is shown
to remain \emph{finite} at the edge of the chaos, contrary to the
varying internal state dynamics \cite{Rokni2007653,druckmann2012neuronal},
for which the settling time is known to diverge \cite{Sompolinsky_Chaos1988_PhysRevLett.61.259}.

\paragraph*{Model and Training Protocol }

Reservoir computing \cite{maass2002real,jaeger2001echo} is a popular
and simple paradigm for training RNN. A network of neurons with random
recurrent connectivity (referred to as the reservoir) is equipped
with readout weights trained to produce a desired output, while keeping
the rest of the connectivity fixed. Such a restricted training rule
implies that training affects reservoir dynamics only via feedback
connections from the output \cite{jaeger2001echo,sussillo2009generating}.
The dynamics (\cite{sussillo2009generating}, \cite{Sompolinsky_Chaos1988_PhysRevLett.61.259,stern2014dyn,Kanaka_PhysRevE.82.011903})
are given by:

\begin{equation}
\dot{x}=-x+Wr+w_{FB}z+w_{in}u\label{eq:FULL_dyn}
\end{equation}
with state $x\in\mathbb{R^{\mathnormal{N}}}$ representing the synaptic
input, and the firing rate given by $r(t)=\phi(x(t))$ where $\phi(x)$
is an element-wise nonlinear function of $x$, commonly set to $\phi(x)=\tanh(x)$.
Output $z=w_{out}^{T}r(t)$ and input $u(t)$ are fed into the network
via weight vectors $w_{FB}\text{ (resp. }w_{in})\in\mathbb{R}^{N}$
with elements i.i.d.. Elements of the connectivity matrix $W\in\mathbb{R}^{N\times N}$
are i.i.d as: $W_{ij}\mathcal{\sim N}(0,g^{2}N^{-1})$ with $g$ being
a gain parameter. 

The goal of the training process is to have the output $z(t)$ approximate
some pre-defined target function $f(t)$. In the reservoir computing
framework training is restricted to modification of the output weights
$w_{out}$. Jaeger \cite{jaeger2001echo} proposed to break the readout-feedback
loop, creating an auxiliary open loop system defined as:

\begin{equation}
\dot{x}=-x+Wr+w_{FB}f+w_{in}u\label{eq:OL_dyn}
\end{equation}
Here the target function $f$, rather than the readout $z$, is injected
via the feedback weights $w_{FB}$. Linear regression on $r$ is used
to find $w_{out}$ so that $z_{OL}=w_{out}^{T}r\approx f$. 

In our case, we assume zero input ($u\equiv0$), and target multiple
fixed points of \eqref{eq:FULL_dyn}, corresponding to $M\ll N$ output
levels $z\in\{A_{1},..A_{M}\}$ with respective solutions $\bar{x}_{1},...\bar{x}_{M}$
and rates $\bar{r}_{1}...\bar{r}_{M}$ which are obtained from the
open loop system \eqref{eq:OL_dyn}. 

\begin{figure*}
\includegraphics[width=17cm]{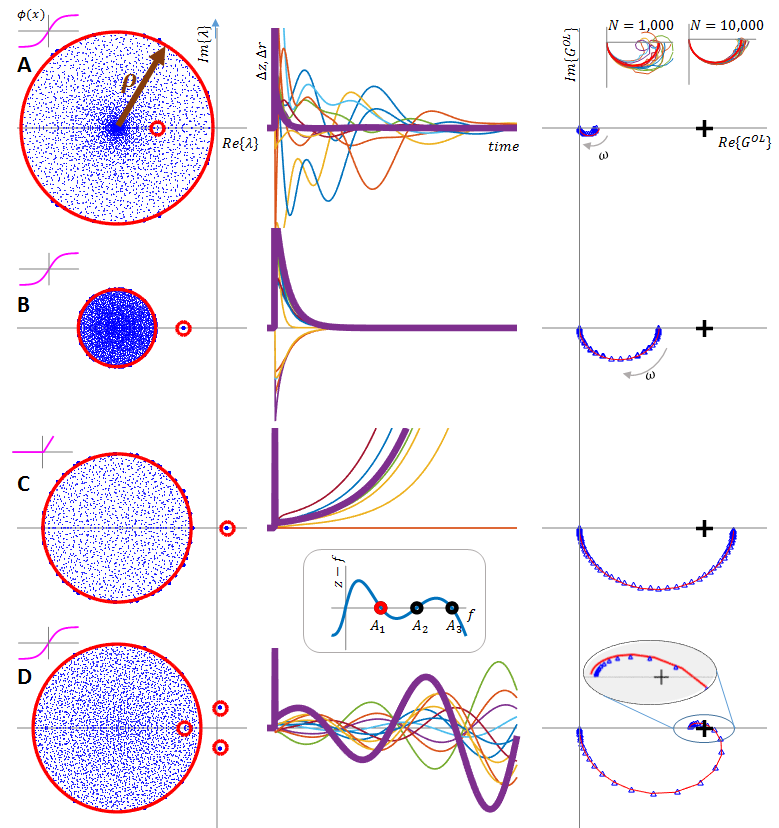}

\caption{\label{TauDrawings} Analysis of a trained RNN is shown for representative
cases compliant with fading memory property ($\rho<1$). \textbf{A}
Internal dynamics are slow compared to network output ($\tau_{net}>\tau_{out}$)
\textbf{B} Opposite case ($\tau_{net}<\tau_{out}$), here the internal
state is dominated by output feedback \textbf{C }Unstable case \textbf{($\tau_{out}<0$)
D }Unstable oscillatory\textbf{ }solution around one of the targets
for $M=3$.\textbf{ Left:} Mean field estimate (red) of closed loop
spectrum compared with a finite size realization (blue dots, $N=3000$).
\textbf{Middle: }Transient response for a $\delta-$like perturbation
is shown for both output (thick line) and for 10 random neurons (thin
lines). \textbf{Right:} MFT estimation (red) of open loop gain is
compared with a finite size realization (blue). The black cross at
$0+1i$ helps visualize the Nyquist criterion. \textbf{Panel A inset:}
Finite size effects (for other cases, where $\rho$ is significantly
smaller than unity finite size effects are small and not shown). \textbf{Parameters}:
Output value was set to $A=1$ for all the cases except \textbf{D}
where $A_{1,2,3}=\{0.5,1.0,1.5\}$ (inset), and $A_{1}$ is analysed.
Nonlinearity $\phi(x)=\tanh(x)$ was used except panel \textbf{C}
for which $\phi(x)=\max(0,x-0.1)$. The connectivity strength scale
$g$ was set to 1.5, 0.5, 1.1 and 1.0 for panels \textbf{A,B,C }and\textbf{
D} respectively.}

\end{figure*}

\paragraph*{Dynamics of a trained network }

A necessary condition for successful training is the fading memory
property \cite{jaeger2001echo} which states that the open loop system
\eqref{eq:OL_dyn} must be globally asymptotically stable for the
training to succeed. Remarkably, asymptotic stability can hold for
suitable drive $f$ %
\footnote{We avoid referring to such a signal as an \emph{input} because in
our study it often refers to a \emph{clamped feedback}. %
} even in systems that are chaotic in the absence of external drive
($f\equiv0$) \cite{yildiz2012re,manjunath2013echo}. In supplemental
material we show that this extended version of fading memory is necessary
even for the FORCE algorithm \cite{sussillo2009generating}, known
for its effectiveness for training intrinsically chaotic networks. 

For a given target $f(t)\equiv A$, fading memory implies that the
open loop system \eqref{eq:OL_dyn} converges to a unique stable state
$\bar{x}$, given by

\begin{equation}
\bar{x}=W\phi(\bar{x})+w_{FB}A\label{eq:FP_equation}
\end{equation}
and that the spectral radius $\rho$ of the linearized open loop dynamics
$WR'$, given by $\rho^{2}=g^{2}<r'^{2}>$ \cite{Ahmadian_PhysRevE.91.012820,Massar_PhysRevE.87.042809},
is smaller than $one$. Here $R'_{ij}=\delta_{ij}r'_{i}$ with $r'=\phi'(\bar{x})=\frac{d\phi}{dx}|_{x=\bar{x}}$
is a diagonal matrix of linearized rate functions, and the average
is taken over neurons.

Importantly, asymptotic stability of the open loop system \eqref{eq:OL_dyn}
does not guarantee stability of the \emph{closed loop} system \eqref{eq:FULL_dyn}.
This can be understood by considering the linearization of the latter:
\begin{equation}
\delta\dot{x}=\left(-I+(W+w_{FB}w_{out}^{T})R'\right)\delta x\label{eq:timeDomainLin}
\end{equation}
For large $N$, the resulting spectrum consists of a disk-like spectral
density region of radius $\rho$ associated with $WR'$ as in the
open loop system and other eigenvalues related to the feedback loop
term $w_{FB}w_{out}^{T}$. We will show that exactly $M$ eigenvalues
correspond to the latter and that their loci can fall either inside
or outside the spectral density disk. Figure \ref{TauDrawings} shows
how these loci determine stability, convergence times and oscillations
for networks that comply with fading memory.

We will derive these eigenvalues of the closed loop system by analyzing
the open loop gain - the response of the open loop output to a small
perturbation in the drive $f=A+\delta f(t)$. In Fourier domain the
state perturbation $X(\omega)$ is given by

\begin{equation}
i\omega X(\omega)=-X(\omega)+WR'X(\omega)+w_{FB}F(\omega),\label{eq:x_smallSignal}
\end{equation}
leading to the open loop gain:

\begin{equation}
G^{OL}(\omega)=Z(\omega|F(\omega)\equiv1)=w_{out}^{T}R'X(\omega|F(\omega)\equiv1).\label{eq:G_OL}
\end{equation}
In the closed loop case $Z(\omega)$ is fed back via $w_{FB}$ and
the gain is given by: 
\begin{equation}
G^{CL}(\omega)=G^{OL}(\omega)(1-G^{OL}(\omega))^{-1}\label{eq:G_CL}
\end{equation}
Poles of \eqref{eq:G_OL} and of \eqref{eq:G_CL} correspond to the
spectrum of linearized versions of \eqref{eq:OL_dyn} and\eqref{eq:FULL_dyn}
respectively. While, in general, all of the $N$ poles can potentially
be modified by closing the loop and transitioning from \eqref{eq:G_OL}
to \eqref{eq:G_CL}, the mean field estimate of $G^{OL}$ which we
now develop is shown to be of an order $M\ll N$, implying that due
to a massive pole-zero cancellation only loci of $M$ eigenvalues
are updated.

We first estimate $G^{OL}(\omega)$ for $N\rightarrow\infty$ for
$M=1$ using second order statistics of $\bar{x}$ and $X$ obtained
from Mean Field Theory. Following the notation in \cite{Kanaka_PhysRevE.82.011903},
we denote the deterministic (independent of $W$) part of the solution
$\bar{x}$ of \eqref{eq:FP_equation} by $\bar{x}^{0}$ and the stochastic
one by $\bar{x}^{1}$. Namely, we have $\bar{x}^{0}=w_{FB}A$ and
$\bar{x}^{1}=W\phi(\bar{x})$ with elements $\bar{x}_{i}^{1}$ distributed
as $\bar{x}_{i}^{1}\sim\mathscr{N}(0,\sigma^{2})$. Variance $\sigma^{2}$
of an individual element of the state vector can be obtained self
consistently, according to:

\begin{equation}
\sigma^{2}=g^{2}\int\mathscr{D}w\int Dy\,\phi^{2}(wA+\sigma y)\label{eq:self_const_LS}
\end{equation}
where $Dy=(\sqrt{2\pi})^{-1}dy\,\exp(-y^{2}/2)$ and $\mathscr{D}w=dw\, p_{w_{FB}}(w)$
correspond to integration with respect to a unity variance Gaussian
measure and to the feedback weight distribution respectively. The
solution $X$ of \eqref{eq:x_smallSignal} is represented similarly
to the state vector $\bar{x}$, but with the stochastic part $X^{1}$
further decomposed into a component fully correlated with $\bar{x}^{1},$
and a component orthogonal to $\bar{x}^{1}$, defined by $X_{\parallel}^{1}=\alpha(\omega)\bar{x}^{1}$
and $<X_{\perp}^{1},\bar{x}^{1}>\equiv0$ respectively. Here and it
what follows we use the notation $N^{-1}a^{T}b=<a,b>$ for self-averaging
quantities. From the equations \eqref{eq:FP_equation} and \eqref{eq:x_smallSignal}
the correlation between $\bar{x}^{1}$ and $X^{1}$ can be expressed
as: 

\begin{multline}
(1+i\omega)<\bar{x}^{1},X_{\parallel}^{1}>=(1+i\omega)<\bar{x}^{1},X^{1}>=\\
<W\bar{r},WR'X>=<W\bar{r},WR'\left(X^{0}+X_{\parallel}^{1}+X_{\perp}^{1}\right)>\label{eq:alpha_Prep}
\end{multline}

Apart from $X_{\perp}^{1}$, this is a self consistent definition
of $\alpha(\omega)$. To ignore $X_{\perp}^{1}$ one argues that the
vectors $\bar{x}^{1}=W\bar{r}$ and $X^{1}=(1+i\omega)^{-1}WR'X$
both result from a product with $W$ and are thus jointly Gaussian.
Orthogonality to $\bar{x}^{1}$, thus renders the vector $X_{\perp}^{1}$\emph{independent}
of $\bar{x}^{1}$, and of all its functions. Consequently, the term
$<W\bar{r},WR'X_{\perp}^{1}>$ vanishes, and realizing that $<Wa,Wb>=g^{2}<a,b>$
 we obtain a self consistency equation for $\alpha(\omega)$:

\begin{equation}
(1+i\omega)\alpha(\omega)=\beta_{0}X^{0}+\beta_{1}\alpha(\omega)\label{eq:betaP_PP}
\end{equation}
with $X^{0}=(1+i\omega)^{-1}w_{FB}$ and 
\begin{equation}
\beta_{0,1}\equiv g^{2}\sigma^{-2}\int\mathscr{D}w\int Dy\;\phi(wA+\sigma y)\phi'(wA+\sigma y)\xi_{0,1}\label{eq:Beta}
\end{equation}
where $\xi_{0}=w$, $\xi_{1}=\sigma y$. 

The readout vector $w_{out}$ in the case of $M=1$ is simply the
vector $\bar{r}$, normalized and scaled by the desired output amplitude:
$w_{out}=A(\bar{r}^{T}\bar{r})^{-2}\bar{r}=N^{-1}g^{2}\sigma^{-2}A\bar{r}$.
Substituting into \eqref{eq:G_OL} yields $G^{OL}(\omega)=g^{2}\sigma^{-2}A<\bar{r}R'X(\omega)>$
and hence:

\begin{equation}
G_{00}(\omega)=<\bar{r}R'X(\omega)>=(1+i\omega)g^{-2}\sigma^{2}\alpha(\omega)\label{eq:G_00}
\end{equation}

\begin{align}
G^{OL}(\omega) & =g^{2}\sigma^{-2}AG_{00}=\frac{A\beta_{0}}{(1-\beta_{1}+i\omega)}\label{eq:G_singleFP}
\end{align}
where the intermediate term $G_{00}$ was defined to facilitate generalization
for $M>1$ below.

Consequently, closed loop system \eqref{eq:FULL_dyn} with gain $G^{CL}$
\eqref{eq:G_CL} has a single (uncanceled) pole located at: 
\begin{equation}
\lambda_{out}=-\left(1-A\beta_{0}-\beta_{1}\right)\label{eq:p_CL}
\end{equation}
this pole corresponds to a single eigenvalue of \eqref{eq:timeDomainLin},
while the rest of its spectrum, corresponding to canceled poles, remains
intact with respect to the open loop disk (Fig. \ref{TauDrawings}A,B,C).

\paragraph*{Robustness and stability of the output}

For a commonly used $\phi(x)=\tanh(x)$ and, more generally for any
sigmoidal activation functions $\phi(x)$ centered at the origin (i.e.
$\phi(0)=\phi''(0)=0$ ), $\lambda^{CL}$ is always negative and the
trained system is thus always \emph{stable}. Conversely, it is always
\emph{unstable} for rectified linear activation function $\phi(x)=\max(0,x-x_{th})$
with positive threshold $x_{th}.$ To check that, one substitutes
integral expressions for $\sigma^{2}$ and for $\beta_{0,1}$ into\eqref{eq:p_CL}
yielding:
\begin{multline}
\lambda_{out}=-\sigma^{-2}g^{2}\int\mathscr{D}wDy\phi(x')\left(\phi(x')-x'\phi'(x')\right)\label{eq:tauIntegralForm}
\end{multline}
where $x'=wA+\sigma y$, and observes that the integrand is always
non-negative (resp. non-positive) for origin-centered sigmoid (resp.
rectified linear) activation function. The situation with all positive,
saturating activation functions \cite{mastrogiuseppe2016intrinsically}
is more complicated and both stable and unstable settings exist.

The pole that was discussed above dictates the settling time constant
$\tau_{out}\equiv-(\lambda_{out})^{-1}$ of a perturbed output. Importantly,
the Maximum Lyapunov Exponent of the system \eqref{eq:FULL_dyn} does
not necessarily coincide with $\lambda_{out}$, but rather with $\max(\lambda^{CL},\rho-1)$.
In particular, for sigmoids mentioned above, $\tau_{out}$ remains
finite even for networks at the edge of the chaos, where, by definition,
the time constant of the internal activity diverges as $\tau_{net}=(1-\rho)^{-1}$
\cite{Sompolinsky_Chaos1988_PhysRevLett.61.259,Ahmadian_PhysRevE.91.012820}.
This possibility of $\tau_{net}\gg\tau_{out}$ is demonstrated in
Figure \ref{TauDrawings}A and can explain the experimental observation
\cite{Rokni2007653,druckmann2012neuronal} of the robustness of functionally
important signals in the presence of highly varying underlying neural
activity.

Validation of the Mean Field Theory by comparison of predicted and
actual spectra is not always meaningful (e.g. Fig \ref{TauDrawings}A).
We thus compare the MFT estimation of $G^{OL}(\omega)$ from equation
\eqref{eq:G_singleFP}, and later \eqref{eq:gainSumOfProj}, with
numerical simulation for finite $N$. Convergence of $G^{OL}(\omega)$
to its MFT estimate is shown Figure \ref{TauDrawings}A (inset), demonstrating
how the ripple in $G^{OL}(\omega)$ vanishes due to the improving
accuracy of pole-zero cancellation as $N$ grows, or equivalently
the subspace $X_{\perp}$becomes unobservable from the output point
of view.

Remarkably, the subspace $X_{\perp}^{1}$ , responsible for this cancellation,
can be used by adaptive algorithms (e.g. FORCE \cite{sussillo2009generating})
for improving the stability of training targets which turn out to
be unstable with a naive LMS readout that we used in this work.

Note that Equation \eqref{eq:G_singleFP} implies that a DC open loop
gain smaller than unity ($G^{OL}(\omega=0)<1$) is a sufficient and
necessary condition for stability of \eqref{eq:FULL_dyn}. This is
not the case for general $M$ as will be shortly shown.

\paragraph*{Multiple Training Targets }

The Least Mean Square readout weight vector in this case is given
by: 
\begin{equation}
w_{out}=N^{-1}\sum_{m=1}^{M}k_{m}\bar{r}_{m}\label{eq:LMS readout}
\end{equation}
where the coefficient vector $k$ is derived from the correlation
matrix of the states $\bar{r}$. The open loop gain around $n-$th
fixed point is hence: 
\begin{equation}
G_{n}^{OL}(\omega)=\sum_{m=1}^{M}k_{m}G_{nm}(\omega).\label{eq:gainSumOfProj}
\end{equation}
with diagonal term $G_{nn}$ calculated as in \eqref{eq:G_00} and
cross terms $G_{nm}(\omega)=<\bar{r}_{m}^{T}R'_{n}X_{n}(\omega)>$
which can be brought to a form:

\begin{equation}
G_{nm}(\omega)=\frac{K_{nm}(i\omega-z_{nm})}{(i\omega-p_{nn})(i\omega-p_{nm})}\label{eq:Gmn_term}
\end{equation}
with $K_{nm}$, $z_{nm}$, $p_{nm}$ and $p_{nn}$ derived in the
supplementary material. Thus we conclude that the local dynamics of
the output of the closed loop system \eqref{eq:G_CL} is governed
by an $M$-th order ODE. This follows from noting that the sum of
Equation \eqref{eq:gainSumOfProj} renders $G_{n}^{OL}(\omega)$ and
$G_{n}^{CL}(\omega)$ $M$-th order rational functions of $\omega$.

Matlab code for the mean field calculation of $G_{OL}(\omega)$ is
provided as supplementary material along with a detailed derivation
of \eqref{eq:Gmn_term}. 

The higher order of $G_{CL}$ in a multiple fixed point setting implies
that the stability condition on the DC gain $G^{OL}(\omega=0)<1$
is no longer sufficient. A counterexample, shown in Fig. \ref{TauDrawings}D,
demonstrates the emergence of complex poles corresponding to unstable
oscillatory behavior. Thus, stability requires evaluation of all $M$
poles of $G^{CL}(\omega)$. Alternatively. the Nyquist criterion \cite{Nyquist_Stab1932,astrom2010feedback_ch9}
can be applied to the open loop system $G^{OL}(\omega)$ avoiding
direct analysis of $G^{CL}(\omega)$. Specifically, stability depends
on whether the curve $G^{OL}(\omega)$ from $-\infty$ to $+\infty$
does not encircle the point $0+1i$ in the complex plane (black crosses
in Figure \ref{TauDrawings})%
\footnote{Due to echo state property, the open loop system is stable, and the
criterion is necessary and sufficient.%
}.

\begin{figure}[H]
\includegraphics[width=8cm]{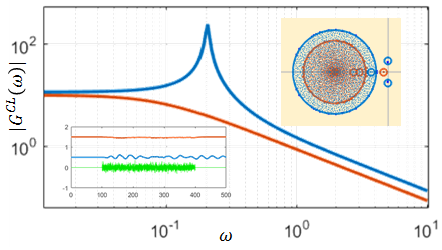}\caption{\label{fig:Validity-of-frequency-selectivity}Network, with settings
of Fig. \ref{TauDrawings}D but $g=0.9$ has stable fixed points at
$A_{1}=0.5$ (blue) $A_{3}=1.5$ (orange). It exhibits frequency selectivity
around the lower fixed point $A_{1}$. At the higher fixed point $A_{3}$
no such selectivity exists. $G^{CL}$ for both cases is shown along
with the spectrum (top inset) and transient response for the same
white noise input (green) delivered through $w_{in}$ to both fixed
points. }
\end{figure}

Importantly, stable resonances may also emerge due to the same mechanisms.
Resonances are characteristic to a specific steady state $z=A_{n}$
of the network, rather than to the network in general. Figure \ref{fig:Validity-of-frequency-selectivity}
demonstrates such a state dependent frequency selectivity in a bi-stable
network. Such selectivity is well known in biological neural circuits
\cite{buzsaki2006rhythms,siegel2015cortical}, and our theory suggests
that it can emerge as an inherent consequence of having multiple steady
states (e.g. fixed points) rather than due to some dedicated frequency
adaptation process. Remarkably, resonance emerges by perturbing through
an arbitrary input $w_{in}$ in \eqref{eq:FULL_dyn}, and not only
through $w_{FB}$ since the resonant eigenvalues shown in figure \ref{fig:Validity-of-frequency-selectivity}
also dictate the slowest timescale of the system as a whole, regardless
of input details.

While no fully analytical treatment for the resonance characteristics
is available, we note that we commonly observed resonance frequencies
in the range of $\omega_{0}\approx0.1-0.5$. Interestingly, Rajan
et al. \cite{Kanaka_PhysRevE.82.011903} predicted an enhanced chaos
suppression by stimuli in a very similar frequency range, indicating
a possible connection between the two phenomena. The supplementary
material contains several bounds on these frequencies, but a full
analysis is beyond the scope of the current work.

In conclusion, we considered high dimensional networks adapted to
produce a desired low dimensional output. The output is being interpreted
here as a firing rate, but can also stand for a stable gene expression
\cite{Ciliberti21082007}, and a variety of other observables \cite{barzel2013universality}.
In all these cases, the network's internal state remains high dimensional
and hard to interpret or investigate directly. The method of combining
mean field approach with system analysis presented here enables predictions
ranging from instability to extreme robustness of the network of interest.
\begin{acknowledgments}
We thank Larry Abbott, Naama Brenner, Lukas Geyrhofer, Vishwa Goudar,
Leonid Mirkin, Daniel Soudry and Merav Stern for their valuable comments.
OB is supported by ERC FP7 CIG 2013-618543 and by Fondation Adelis.

Supplemental Material can be found at: \href{http://barak.net.technion.ac.il/publications/}{http://barak.net.technion.ac.il/publications/}.
\end{acknowledgments}
\bibliographystyle{\string"C:/Program Files (x86)/MiKTeX 2.9/bibtex/bst/revtex4/apsrev_noURL\string"}
\bibliography{arivkindNeuroBib}

\end{document}